\documentclass [prd,eqsecnum,aps, twocolumn, noshowpacs]{revtex4}
\usepackage{epsf,graphics,graphicx}
\usepackage{amsmath}
\usepackage{amssymb,latexsym,mathrsfs}
\usepackage{hyperref}

\def\bea{\begin{eqnarray}}
\def\eea{\end{eqnarray}}
\def\ba{\begin{array}}
\def\ea{\end{array}}

\def\beq{\begin{equation}}
\def\eeq{\end{equation}}

\begin{document}
%\title{Constraining near de-Sitter inflation by WMAP 7 years data using Hamilton Jacobi formulation}
%\title{Constraining a general class of exponential inflation by WMAP using Hamilton Jacobi}
\title{Confronting quasi-exponential inflation with WMAP seven}
%\title{Constraining quasi-exponential inflation by WMAP seven}

\author{Barun Kumar Pal$^{1}$\footnote{Electronic address: {barunp1985@rediffmail.com}},
 ${}^{}$ Supratik Pal$^{1, 2}$\footnote{Electronic address: {pal@th.physik.uni-bonn.de}} ${}^{}$
and B. Basu$^{1}$\footnote{Electronic address:
{banasri@isical.ac.in}} ${}^{}$} \affiliation{$^1$Physics and
Applied Mathematics Unit, Indian Statistical Institute, 203 B.T.
Road, Kolkata 700 108,
India \\
$^2$Bethe Center for Theoretical Physics and Physikalisches
Institut der Universit\"{a}t Bonn, Nussallee 12, 53115 Bonn,
Germany}

\vspace{1in}

\begin{abstract}
We confront quasi-exponential models of inflation with WMAP seven
years dataset using Hamilton Jacobi formalism. With a phenomenological Hubble parameter, representing
quasi exponential inflation, we develop the formalism and
subject the analysis to confrontation with WMAP seven using the
publicly available code CAMB. The observable parameters are found
to fair extremely well with WMAP seven. We also obtain a ratio of
tensor to scalar amplitudes which may be detectable in PLANCK.
\end{abstract}

%Early Universe, Cosmology, Origin and formation of the Universe, Gravitational waves, Background radiation, Observational cosmology
%\pacs{98.70.Vc; 98.80.Es; 98.80.Bp}

\vspace{1in} \maketitle
%%%%%%%%%%%%%%%%%%%%%%%%%%%%%%%%%%%%%%%%%%%%%%%%%%%%%%%%%%%%%%%%%%%%%%%%%%%%%%%%%%%%%%%%%%%%%%%%%%%%%%%%%%%%%%%%%%%%%%%%%%%%%%%%%%%%%%%%%%%%%%%%%%%%%%%%
%%%%%%%%%%%%%%%%%%%%%%%%%%%%%%%%%%%%%%%%%%%%%%%%%%%%%%%%%%%%%%%%%%%%%%%%%%%%%%%%%%%%%%%%%%%%%%%%%%%%%%%%%%%%%%%%%%%%%%%%%%%%%%%%%%%%%%%%%%%%%%%%%%%%%%%%
%\tableofcontents

\section{Introduction}
Present-day cosmology is inflowing into an era where it is becoming more and more possible to constrain the models of the early  universe by precise data coming from highly
 sophisticated observational probes like WMAP \cite{wmap}, PLANCK \cite{planck}, SDSS \cite{sdss}, ACBAR \cite{acbar}, QUaD \cite{quad}.
Such observations are gradually leading theoretical cosmology towards the details of physics at very high energies, and the possibility of testing some of 
the speculative ideas of recent years. Inflation -- the most fascinating  among them  -- was first proposed back in 1981 by Guth \cite{guth}, and has since been developed by a
 wide variety of probable models \cite{2,3,4,5,7,8,9,10}. However, the inflationary scenario is more like a paradigm than a theory, since
a specific compelling model is yet to be separated out from the spectrum of possible models and alternatives.

So far the most appealing prediction of inflationary paradigm was
that it may generate spectra of both density perturbations  and
gravitational waves. Before the detection of CMB anisotropies by
COBE \cite{cobe}, cosmological observations had limited range of
scales to access, and it was sufficient to predict from an
inflationary scenario  a scale-invariant spectrum of density
perturbations and a negligible amplitude of gravitational waves.
However, since COBE \cite{cobe} and, of late, WMAP \cite{wmap} came
forward,  the spectra are now well-constrained over a wide range
of scales ranging from 1 {\rm Mpc} upto 10,000 {\rm Mpc}. So the
inflationary predictions should now be very precise  to
incorporate latest observations, say, from WMAP seven year run.
The ongoing satellite mission PLANCK \cite{planck}
will restrict the theoretical predictions further. It promises to
survey the ratio of the tensor to scalar amplitudes $r$ upto the
order of $10^{-2}$ \cite{colombo} (orders below WMAP \cite{wmap} predictions $r<0.36$ at 95\%
Confidence Level [C.L.] \cite{komatsu}) so as to comment more precisely on primordial
gravitational waves and the energy scale of inflation. PLANCK may
even discriminate single field inflationary models from multi-field models by detecting
(provided $f_{NL}\geq 5$) large {\it non gaussianity}. So this is
high time to confront different class of inflationary models with
latest data and forthcoming predictions.

The usual technique used in the inflationary scenario to solve the dynamical equations is the {\it slow roll approximations} \cite{liddle}. But it is not the only possibility for 
successfully implementing models of inflation and solutions outside the {\it slow roll approximations} have been found \cite{wands}. To incorporate all the models irrespective of 
slow roll approximations {\it Hamilton Jacobi} formalism \cite{grishchuk,muslimov,salopek} turned out to be very
useful. The formulation is imitative by considering the {\it inflaton} field itself to be the evolution parameter. The key advantage of this formalism is that here we only need the 
{\it Hubble parameter} $H(\phi)$ to be specified rather than the inflaton potential $V(\phi)$. Since $H$ is a geometric quantity,
unlike $V$, inflation is more naturally described in this language \cite{muslimov,salopek}. Further, being first order in nature, these equations are easily tractable to explore the 
underlying physics. Once $H(\phi)$ has been specified, one can, in principle, derive a relation between $\phi$ and $t$ which will
enable him to get hold of $H(t)$ and the scale factor $a(t)$ therefrom. Therefore {\it Hamilton Jacobi} formalism provides us with a straightforward way of exploring inflationary 
scenario and related observational aspects.

In this article,  we intend to confront quasi-exponential inflationary models with WMAP seven years data \cite{wmap} with a 
phenomenological Hubble parameter following  {\it Hamilton Jacobi} formalism. The absence of time dependence in the Hubble parameter 
produces de-Sitter inflation which was very appealing from theoretical point of view but its acceptability is more or less limited 
considering present day observations. So, certain deviation from an exact exponential inflation turns out to be a good move so as to go 
along with latest as well as forthcoming data. In general these models are called quasi-exponential
inflation. Our primary intention here is to confront this quasi-exponential inflation with WMAP seven using the publicly 
available code CAMB \cite{camb}. Nevertheless, as it will turn out, the analysis also predicts a detectable tensor to scalar ratio so 
as to reflect significant features of quasi-exponential inflation in forthcoming PLANCK \cite{planck} data sets  as well. 
Thus, PLANCK may directly verify (or put further
constraint to) our analysis by detecting (or pushing further below the upper bound of) gravitational waves.

%%%%%%%%%%%%%%%%%%%%%%%%%%%%%%%%%%%%%%%%%%%%%%%%%%%%%%%%%%%%%%%%%%%%%%%%%%%%%%%%%%%%%%%%%%%%%%%%%%%%%%%%%%%%%%%%%%%%%%%%%%%%%%%%%%%%%%%%%%%%%%%%%%%%
%%%%%%%%%%%%%%%%%%%%%%%%%%%%%%%%%%%%%%%%%%%%%%%%%%%%%%%%%%%%%%%%%%%%%%%%%%%%%%%%%%%%%%%%%%%%%%%%%%%%%%%%%%%%%%%%%%%%%%%%%%%%%%%%%%%%%%%%%%%%%%%%%%%%%%%%
\section{Modeling quasi-exponential inflation by Hamilton Jacobi}
The {\it Hamilton Jacobi} formalism allows us to express the
Friedmann equation in terms of a first order differential equation \cite{grishchuk,muslimov,salopek,kinney}
\beq\label{hamilton}
 \left[H^{'}(\phi)\right]^2 -\frac{3}{2M_P^2}
  H(\phi)^2=-\frac{1}{2M_P^4}V(\phi)
\eeq and the evolution of the scalar field \beq\label{phidot}
\dot{\phi}=-2M_P^2 H'(\phi)
\eeq
where a prime denotes a derivative with respect to the scalar field $\phi$ and a dot a time derivative. The above two equations govern the inflationary
dynamics in Hamilton Jacobi formalism. The shape of the associated potential can be obtained by rearranging terms of Eqn.(\ref{hamilton}) to give
 \beq\label{potential}
 V(\phi)=3M_P^2H^2(\phi)\left[1-\frac{1}{3}\epsilon_H\right]
 \eeq
 where $\epsilon_H$ has been defined as
\beq\label{epsilon} \epsilon_H=2M_P^2\left(\frac{H^{'}}{H}
\right)^2 \eeq We further have \beq\label{adot}
\frac{\ddot{a}}{a}=H(\phi)^2\left[1-\epsilon_H\right] \eeq
 So  $\epsilon_H < 1$ implies accelerated expansion. It is
 customary to define another parameter by
 \begin{equation}\label{eta}
 \eta_H=2M_P^2~ \frac{H^{''}}{H}
\end{equation}
It is worthwhile to mention here that the parameters $\epsilon_H$ and $\eta_H$ are not the usual  slow roll parameters. But in the
slow roll limit  $\epsilon_H \rightarrow\epsilon$ and $\eta_H\rightarrow\eta-\epsilon$ \cite{liddle}, $\epsilon$ and
$\eta$  being usual {\it slow roll} parameters.

Let us consider a phenomenological Hubble parameter representing quasi exponential inflation
 \beq\label{hubble}
 H(\phi)=H_{inf} \exp\left[\frac{\frac{\phi}{M_P}}{p(1+\frac{\phi}{M_P})}\right]
\eeq where $p$ is a dimensionless parameter and $H_{inf}$ is a
constant having dimension of Planck Mass. The value of the
constants can be fixed from the conditions for successful
inflation and the observational bounds.
As it will turn out in subsequent analysis this Hubble parameter can indeed be cast into a form of quasi exponential
inflation for some choice in the parameter space.

The two parameters $\epsilon_H$ and $\eta_H$  in the Hamilton
Jacobi formalism take the form
\begin{equation}\label{slowroll}
\epsilon_H = \frac{2}{p^2(1+\frac{\phi}{M_P})^4}, ~~~ \eta_H
=-\frac{2\left(-1+2p+2p\frac{\phi}{M_P}\right)}{p^2(1+\frac{\phi}{M_P})^4}
\end{equation}
\begin{figure}%[htb]
  \centerline{\includegraphics[width=8.5cm, height=5.5cm]{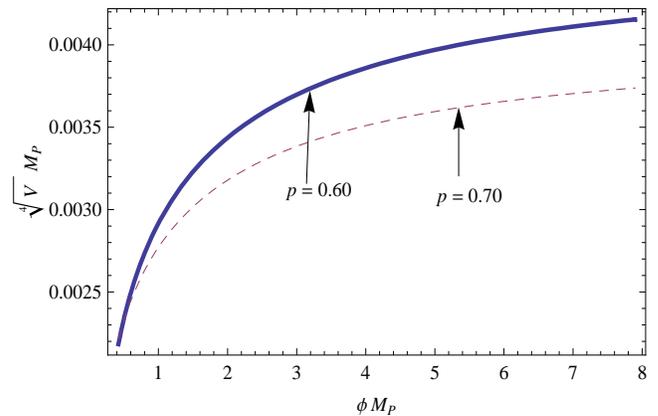}}
  \caption{\label{energy} Variation of the fourth root of the potential with the scalar field}
\end{figure}

Now, the condition for inflation may  be put forward using
equation of state parameter which has the form
\beq\label{eqnstate}
\omega(\phi)\equiv\frac{P(\phi)}{\rho(\phi)}=\frac{4}{3p^2(1+\frac{\phi}{M_P})^4}-1
\eeq
 \begin{figure}%[htb]
 \centerline{\includegraphics[width=8.5cm, height=5.50cm]{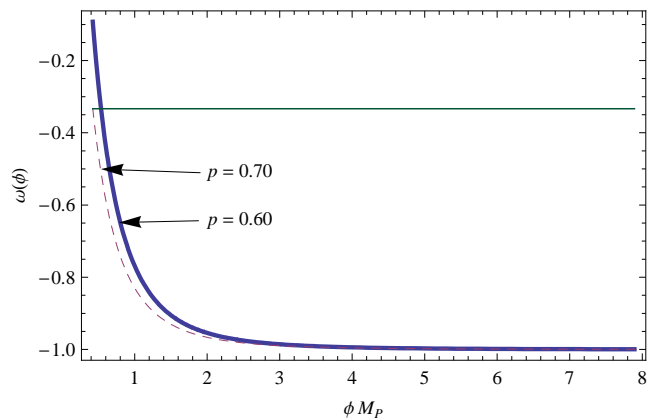}}
 \caption{\label{eos} Variation of the equation of state parameter with the scalar field $\phi$, the horizontal line represents $\omega=-\frac{1}{3}$}
\end{figure}

From the constraint $\omega < -\frac{1}{3}$ during inflation, we
get \beq\label{plower} |p|>\frac{\sqrt{2}}{(1+\frac{\phi}{M_P})^2}
\eeq So for almost any value of $p$ inflation is achievable.
However, at the end of inflation $\omega \geq -\frac{1}{3}$
allows us to estimate an upper bound for $p$ as \beq\label{pupper}
 |p|\leq\frac{\sqrt{2}}{(1+\frac{\phi_{end}}{M_P})^2}
\eeq where  $\phi_{end}$   is the value of the inflaton field at
the end of inflation. In order to implement a successful model of
inflation both the restrictions on $\omega$ should be satisfied.
First note that if $p$ is negative then from Eqn.(\ref{phidot})
and Eqn.(\ref{hubble}) we have $\dot{\phi}>0$, i.e. $\phi(t)$
increases with time which incorporates the so called
 {\it graceful exit problem} \cite{powerlaw,pow1}. We discard the values of $p$ greater or equal to $\sqrt{2}$ as well, since $p\geq\sqrt{2}$
would imply $\epsilon_H<1$ for any value of $\phi$ giving rise to the same problem. As a result the feasible range for $p$ would be:
$0<p<\sqrt{2}$. Further, for sufficient inflation we also need $|\eta_H|<1$ during inflation. And violation of that condition
after inflation drags the inflaton towards its potential minima quickly. But if $p<0.586$ then $|\eta_H|$ will always be less than
one. Imposing this  condition further restricts the range of the otherwise free parameter $p$ within $0.586<p<\sqrt{2}$. 
We shall use a representative value for the parameter $p$ within the above range later on in this paper while confronting CAMB \cite{camb} 
outputs with WMAP seven \cite{wmap}.

Now to derive the expression for the scale factor we shall make
use of the relation
 \bea\label{sc}
 \dot{\phi}\frac{a'(\phi)}{a(\phi)}&=& H(\phi)
 \eea
Plugging Eqn.(\ref{hubble}) into Eqn.(\ref{phidot})  and using
Eqn.(\ref{sc}) we have \beq\label{aphi}
a(\phi)=a_e\exp\left[-\frac{p(1+\frac{\phi}{M_P})^3}{6}\right]
\eeq where $a_e\equiv a(\phi_{end})
\exp\left[-\frac{p(1+\frac{\phi_{end}}{M_P})^3}{6}\right]$,
$a(\phi_{end})$ being the scale factor at the end of inflation.

 \begin{table}%[htb]
\begin{tabular}{|c|c|c|c|c|c|c|}
\hline $p$  & $\epsilon_H<1$ &$|\eta_H|<1$ & $\phi_{end}$ &
$\phi_{in}$ & N & $V(\phi_{in})^{1/4}$
 \\
&$\phi\geq M_P$ &$\phi\geq M_P$& $M_P$ & $M_P$&&$10^{16}$ GeV   \\
\hline
0.60 & 0.535 & 0.385& 0.535 &7.260 & 56&1.005  \\
& & & &  7.894 & 70&1.012\\
\hline
0.70 & 0.421 & 0.414& 0.421 &6.845 & 56&0.901  \\
& & & & 7.448&70&0.907\\
%\hline
%0.80 & 0.329574 & 0.40553& 0.329574 &6.50282 & 56  \\
%& & & &7.07916&70 \\
\hline
\end{tabular}
\caption{Different parameters for different values of $p$ within
its allowed range} \label{tab1}
\end{table}
The primary quantities related to inflation have been summarized in Table \ref{tab1}. We see
that $|\eta_H| \approx 1$ after the end of inflation, and so {\it
slow roll} would be a very good approximation throughout the
inflationary period, though we do not utterly require them in this
formalism \cite{kinney}.
%%%%%%%%%%%%%%%%%%%%%%%%%%%%%%%%%%%%%%%%%%%%%%%%%%%%%%%%%%%%%%%%%%%%%%%%%%%%%%%%%%%%%%%%%%%%%%%%%%%%%%%%%%%%%%%%%%%%%%%%%%%%%%%%%%%%%%%%%%%%%%%%%%%%%%%%
%%%%%%%%%%%%%%%%%%%%%%%%%%%%%%%%%%%%%%%%%%%%%%%%%%%%%%%%%%%%%%%%%%%%%%%%%%%%%%%%%%%%%%%%%%%%%%%%%%%%%%%%%%%%%%%%%%%%%%%%%%%%%%%%%%%%%%%%%%%%%%%%%%%%%%%%

That the above scale factor indeed represents  quasi-exponential
inflation will transpire from the following analysis. We first
expand $H(\phi)$ into the power series of $\phi$
\beq\label{approxh} H(\phi)=
H_1\left[1-\frac{1}{p(1+\frac{\phi}{M_P})}+\frac{1}{2!}\frac{1}{p^2(1+\frac{\phi}{M_P})^2}-...\right]
\eeq where we have defined $H_1\equiv
H_{inf}\exp\left[\frac{1}{p}\right]$. The above expansion, at next
to leading order, gives rise to
\beq\label{phit}
\phi\approx M_P\left[\left(\frac{6H_1(t-t_e)}{p}\right)^{1/3}-1\right]
\eeq

Here $t_e\equiv \frac{6H_1}{p}
t_{end} +(1+\frac{\phi_{end}}{M_P})^3$ and $t_{end}$ corresponds
to the time at the end of inflation.

 The combination of Eqn.(\ref{aphi}) and Eqn.(\ref{phit}) gives the scale factor as a function of cosmic time $t$
\beq\label{at}
a(t)\approx a_e\exp\left[H_1(t-t_e)\right]
\eeq
So the expression for the conformal time turns out to be
\beq\label{conformal}
\eta\approx-\frac{1}{H_1a(t)}
\eeq
Thus, our
analysis indeed deals with quasi-exponential inflation. The higher order
terms in the expansion (\ref{approxh}) will, in principle, measure
further corrections to the scale factor but the analytical
solutions are neither always obtainable nor utterly required.
Rather, one can directly confront the observable parameters with
WMAP seven in order to constrain quasi-exponential inflation, as
done in the rest of the article.

%%%%%%%%%%%%%%%%%%%%%%%%%%%%%%%%%%%%%%%%%%%%%%%%%%%%%%%%%%%%%%%%%%%%%%%%%%%%%%%%%%%%%%%%%%%%%%%%%%%%%%%%%%%%%%%%%%%%%%%%%%%%%%%%%%%%%%%%%%%%%%%%%%%%%%%%
%%%%%%%%%%%%%%%%%%%%%%%%%%%%%%%%%%%%%%%%%%%%%%%%%%%%%%%%%%%%%%%%%%%%%%%%%%%%%%%%%%%%%%%%%%%%%%%%%%%%%%%%%%%%%%%%%%%%%%%%%%%%%%%%%%%%%%%%%%%%%%%%%%%%%%%%
\section{Perturbations and observable parameters}

%%%%%%%%%%%%%%%%%%%%%%%%%%%%%%%%%%%%%%%%%%%%%%%%%%%%%%%%%%%%%%%%%%%%%%%%%%%%%%%%%%%%%%%%%%%%%%%%%%%%%%%%%%%%%%%%%%%%%%%%%%%%%%%%%%%%%%%%%%%%%%%%%%%%%%%
\subsection{Scalar curvature perturbation}
The $k^{th}$ Fourier mode of the comoving curvature perturbation,
obtained by solving Mukhanov-Sasaki equation
%\begin{equation}\label{ms}
%v_k^{\prime\prime}+\left(k^2-\frac{z^{\prime\prime}}{z}\right)v_k=0
%\end{equation}
%where the field $v$ is related to the comoving curvature perturbation ${\cal R}$ by $v_k=-z{\cal R}_k$ and $z=\frac{a \phi^\prime}{{\cal{H}}}$ and a prime denotes a derivative with respect to conformal time. After some simple calculations the Mukhanov variable $z$ turns out to be
%\bea\label{z}
%z&=&\frac{2M_P}{p H_1\eta\left[\frac{6}{p}\ln\left(H_1a_e|\eta|\right)\right]^{2/3}}
%\eea
%solving the Mukhanov-Sasaki Eqn.(\ref{ms})
in the  {\it slow-roll} limit
%i.e. $\frac{z^{\prime\prime}}{z} \approx \frac{a^{\prime\prime}}{a}$
and adopting standard Bunch-Davies \cite{bd} vacuum as initial
condition, is approximately given by
\beq\label{comcur}
{\cal R}_k\approx-\frac{p H_1\eta\left[\frac{6}{p}\ln\left(H_1a_e|\eta|\right)\right]^{2/3}}{2M_P}
\frac{\exp[-i k\eta]}{\sqrt{2k}}\left(1-\frac{i}{k\eta}\right)
\eeq
So the dimensionless spectrum for the comoving curvature
perturbation is,
\beq\label{scalarspectrum}%|_{k=aH}
P_{\cal R}(k)=\frac{p^2
H^2_1}{16M_P^2\pi^2}\left(1+k^2\eta^2\right)\left[\frac{6}{p}\ln\left(H_1a_e|\eta|\right)\right]^{4/3}
\eeq
 Now to evaluate the spectrum at horizon exit $k=aH$,  we notice that
 \bea\label{hc}
 1+k^2\eta^2
%&=&1+a^2H^2\eta^2\nonumber\\
% &=&1+\frac{H^2}{H^2_1}\nonumber\\
 %&=&1+\exp\left[-\frac{2}{p(1+\frac{\phi}{M_P})}\right] \nonumber\\
 %&\approx&1+1-\frac{2}{p(1+\frac{\phi}{M_P})}\nonumber\\
 &=&2-\frac{2}{\bigg[6H_1p^2(t_e-t)\bigg]^{\frac{1}{3}}}
 \eea
 To arrive at Eqn.(\ref{hc}) we have used Eqn.(\ref{conformal}) and Eqn.(\ref{phit}) respectively.
This reflects the features of quasi-exponential behavior where the
effect of the evolution of the scalar field has been directly
taken into  considerations, without using $k=-\eta^{-1}$
\textit{a priori}. We shall, of course, use the relation
$k=-\eta^{-1}$ finally while estimating parameters. Loosely
speaking, the argument is analogous to putting a specific value
for a variable after integration (which is more accurate) rather
than putting it before (which may be a good approximation at the
best). Somewhat similar treatment, though from a different physical
perspective, has been employed in \cite{hiranya}.

 Therefore the power spectrum for ${\cal R}_k$ evaluated at horizon crossing is now given by
 \beq\label{scalarspectrum1}
P_{\cal R}(k)|_{k=aH}=\frac{p^2
H^2_1}{8\pi^2M_P^2}\left(\left[\frac{6A(k)}{p}\right]^{4/3}-\frac{6A(k)}{p^2}\right)
\eeq where we have defined $A(k)\equiv\ln\left(H_1a_e k^{-1}\right)$. Also at horizon
crossing $d\ln k=H_1dt$, so the scalar spectral index looks
\beq\label{spectral}
n_S(k)=1-\frac{\frac{4}{3}\left[\frac{6}{p}\right]^{4/3}A(k)^{1/3}-\frac{6}{p^2}}{\left[\frac{6A(k)}{p}\right]^{4/3}-\frac{6A(k)}{p^2}}
\eeq and its running \beq\label{run}
n'_S(k)=-\frac{\frac{4}{3}\left(\frac{6}{p}\right)^{8/3}A(k)^{2/3}-\frac{20}{9}\frac{6^{7/3}}{p^{10/3}}A(k)^{1/3}+\frac{36}{p^4}}
{\left(\left[\frac{6A(k)}{p}\right]^{4/3}-\frac{6A(k)}{p^2}\right)^2}
\eeq
Here note that we would get $n_S=1-\frac{4}{3A(k)}$ and $n'_S=-\frac{4}{3A(k)^2}$ the
expressions for the scalar spectral index and its running
respectively if we had put $k=-\eta^{-1}$ directly into
Eqn.(\ref{scalarspectrum}).

%%%%%%%%%%%%%%%%%%%%%%%%%%%%%%%%%%%%%%%%%%%%%%%%%%%%%%%%%%%%%%%%%%%%%%%%%%%%%%%%%%%%%%%%%%%%%%%%%%%%%%%%%%%%%%%%%%%%%%%%%%%%%%%%%%%%%%%%%%%%%%%%%%%%%%%%
\subsection{Tensor perturbation}

The power spectrum for the tensor modes representing primordial
gravitational waves, obtained in a similar way as before, is given
by 
\beq\label{pt}
P_{T}(k)\equiv2P_h(k)=\frac{H_1^2}{\pi^2M_P^2}\left(1+k^2\eta^2\right)
\eeq 
Now using Eqn.(\ref{hc}) we evaluate Eqn.(\ref{pt}) at horizon
crossing \beq\label{pt1}
P_{T}(k)|_{k=aH}=\frac{2H_1^2}{\pi^2M_P^2}\left[1-\frac{1}{[6p^2A(k)]^{1/3}}\right]
\eeq Corresponding spectral index can be derived immediately from
Eqn.(\ref{pt1}) as \beq\label{nt}
n_{T}(k)=-\frac{1}{3}\frac{1}{(6p^2)^{1/3}A(k)^{4/3}-A(k)} \eeq
and the running of the tensor spectral index is given by
\beq\label{ntr}
n^{'}_{T}(k)=-\frac{1}{9}\frac{4[6p^2A(k)]^{1/3}-3}{\bigg[(6p^2)^{1/3}A(k)^{4/3}-A(k)\bigg]^2}
\eeq Here tensor spectral index $n_T=0$ and its running $n'_T=0$, the
results that we would obtain by the direct substitution of
$k=-\eta^{-1}$  into Eqn.(\ref{pt}).
%%%%%%%%%%%%%%%%%%%%%%%%%%%%%%%%%%%%%%%%%%%%%%%%%%%%%%%%%%%%%%%%%%%%%%%%%%%%%%%%%%%%%%%%%%%%%%%%%%%%%%%%%%%%%%%%%%%%%%%%%%%%%%%%%%%%%%%%%%%%%%%%%%%%%%%%

We are now in a position of dealing with one of the most important
observable parameters, namely, the ratio of tensor to scalar
amplitudes. In the present context we have found
\begin{eqnarray}\label{r}
 r %&\equiv& \frac{P_{T}|_{k=aH}}{P_{\cal R}|_{k=aH}}\nonumber\\
 &=& \frac{16}{[6A(k)]^{4/3} p^{2/3}}
\end{eqnarray}
In the next section we shall comment on its numerical estimates.

The consistency relation can be obtained by combining
Eqn.(\ref{nt}) and Eqn.(\ref{r}), which in this case turns out to
be
\beq\label{rnt} r=-8n_T\left(1-\frac{r^{1/4}}{2p^{1/2}}\right)
\eeq
The result is slightly different from the known relation $r=-8n_T$. In the literature
there are other techniques for obtaining a modified consistency relation. Such a modified
consistency relation can be found in any analysis where higher order terms in the expansion
of slow roll parameters are taken into account \cite{higher}. The consistency relation is also
modified in the context of brane inflation \cite{bean,hiranya} and non-standard
models of inflation \cite{mukhanov} where generalized propagation speed (less than one) of the scalar
field fluctuations relative to the homogeneous background have been considered. Further deviation from the usual consistency
relation can be found in \cite{parker} where {\it tensor to scalar ratio} has been shown to be a function of
{\it tensor spectral index, scalar spectral index} and {\it running of the tensor spectral index}. Our approach is somewhat
similar to these.
%%%%%%%%%%%%%%%%%%%%%%%%%%%%%%%%%%%%%%%%%%%%%%%%%%%%%%%%%%%%%%%%%%%%%%%%%%%%%%%%%%%%%%%%%%%%%%%%%%%%%%%%%%%%%%%%%%%%%%%%%%%%%%%%%%%%%%%%%%%%%%%%%%%%%%%%%%
%%%%%%%%%%%%%%%%%%%%%%%%%%%%%%%%%%%%%%%%%%%%%%%%%%%%%%%%%%%%%%%%%%%%%%%%%%%%%%%%%%%%%%%%%%%%%%%%%%%%%%%%%%%%%%%%%%%%%%%%%%%%%%%%%%%%%%%%%%%%%%%%%%%%%%%%%%
\section{Confrontation with WMAP seven}

%%%%%%%%%%%%%%%%%%%%%%%%%%%%%%%%%%%%%%%%%%%%%%%%%%%%%%%%%%%%%%%%%%%%%%%%%%%%%%%%%%%%%%%%%%%%%%%%%%%%%%%%%%%%%%%%%%%%%%%%%%%%%%%%%%%%%%%%%%%%%%%%%%%%%%%%%
\subsection{Direct numerical estimation}

In Table \ref{tab2} we have estimated the observable parameters
from the first principle of the theory of fluctuation as derived
in the previous section, for two sets of values of $p$ within its allowed range, $0.586<p<\sqrt{2}$. For
estimation we have taken $H_{inf}=2.27\times 10^{-6}~{\rm M_P}$,
$a_e=7.5\times10^{-31}$ and set the pivot scale at $k_0=0.002~{\rm
Mpc}^{-1}$.
\begin{table}%[htb]
\begin{tabular}{|c|c|c|c|c|c|c|c|}
\hline $p$ & $P_{\cal R}^{1/2}$ &$n_S $& $n_T $& $n_S'$&$n_T' $& r
 \\
& $10^{-5}$ & &$10^{-3}$ &$10^{-4}$&$10^{-5} $& $10^{-2}$\\
 \hline
0.60&4.912&0.9746&-1.516&-4.253&-3.853&0.967\\
\hline
0.70&4.114&0.9747&-1.343&-4.334&-3.403&0.878\\
\hline
\end{tabular}
\caption{Observable quantities as obtained from the theory of
fluctuations}\label{tab2}
\end{table}
Table \ref{tab2}  ~reveals that the observable parameters as
derived from our analysis are in excellent agreement with the current
observations as given by WMAP seven years data for
$\Lambda$CDM background \cite{wmap}.

Nevertheless, present day observations are eagerly waiting to
detect primordial gravitational waves. The current observational
bound on the ratio of tensor to scalar amplitudes as
given by WMAP is $r<0.36$ at 95\% C.L. \cite{komatsu}. With the
ongoing satellite experiment PLANCK
 promising to detect it down to the order of $10^{-2}$ \cite{colombo} in near future, this parameter is now playing a pivotal role (along with
non-gaussianity) to discriminate among different models \cite{vega}. The
numerical estimation  reveals that the parameter $r$ is of the
order $10^{-2}$ for quasi-exponential inflation. Thus, PLANCK
\cite{planck} may directly verify (or put further constraint
to) our analysis by detecting (or pushing further below the upper
bound of) gravitational waves.

%%%%%%%%%%%%%%%%%%%%%%%%%%%%%%%%%%%%%%%%%%%%%%%%%%%%%%%%%%%%%%%%%%%%%%%%%%%%%%%%%%%%%%%%%%%%%%%%%%%%%%%%%%%%%

\subsection{CAMB output and comparison}

In what follows we shall make use of the  publicly available code CAMB
\cite{camb} in order to confront our results directly with
observational data. For CAMB, the Eqn.(\ref{scalarspectrum1}) has
been set as initial power spectrum and the values of the initial
parameters associated with inflation are taken from the
Table \ref{tab2} for $p=0.60$ within its allowed range $0.586<p<\sqrt{2}$. Also WMAP seven years dataset
for $\Lambda$CDM background has been used in CAMB to
obtain matter power spectrum and CMB angular power spectrum. We have set the pivot scale at $k_0=0.002~{\rm Mpc}^{-1}$.

Table \ref{tab3} shows input from the WMAP seven years dataset
 for $\Lambda$CDM background.

\begin{table}%[htb]
\begin{tabular}{|c|c|c|c|c|c|c|c|c|c|}
\hline $H_0$ & $\tau_{Reion}$ &$\Omega_b h^2$& $\Omega_c h^2 $&
$T_{CMB}$
 \\
km/sec/MPc& & && K\\
 \hline
71.0&0.09&0.0226&0.1119&2.725\\
\hline
\end{tabular}
\caption{Input parameters}\label{tab3}
\end{table}
Table \ref{tab4} shows the output obtained from CAMB, which is in
fine concord with WMAP seven years data. The results obtained here are for a representative value of the parameter $p=0.6$ within its allowed range.
\begin{table}%[htb]
\begin{tabular}{|c|c|c|c|c|c|c|c|c|c|}
\hline $t_0$ & $z_{Reion}$
&$\Omega_m$&$\Omega_{\Lambda}$&$\Omega_k$&$\eta_{Rec}$& $\eta_0$
 \\
Gyr& & && &Mpc & Mpc\\
 \hline
13.708&10.692&0.2669&0.7331&0.0&285.15&14347.5\\
\hline
\end{tabular}
\caption{Different physical quantities as obtained from
CAMB}\label{tab4}
\end{table}

\begin{figure}%[htb]
 \centerline{\includegraphics[width=8.5cm, height=5.5cm]{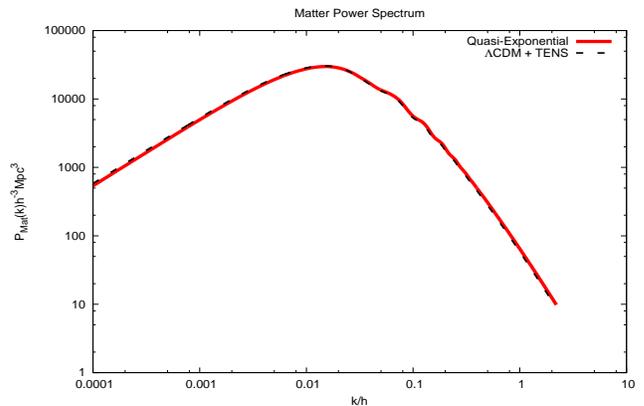}}
 \caption{\label{matpow} Variation of matter power spectrum $P_{Mat}(k)$
  with  $k/h$ in logarithmic scales for quasi-exponential inflation and the best fit spectra of WMAP7 for $\Lambda$CDM + TENS}
\end{figure}
The curvature perturbation is generated due to the fluctuations in
the {\it inflaton} and remains almost constant on the {\it super
Hubble} scales. Long after the end of inflation it makes horizon
reentry creating matter density fluctuations through the
gravitational attraction of the potential wells. These matter
density fluctuations grew with time forming the structure in the
universe. So the measurement of the matter power spectrum is very
crucial as it is directly related to the formation of structure.
In  Fig.\ref{matpow} the CAMB output for the variation of the
spectrum of the matter density fluctuations with the scale for 
quasi-exponential inflation and the best fit spectra of WMAP7 for $\Lambda$CDM + TENS \cite{cdmtens} have
been shown and it represents true behavior indeed \cite{sdss}.

\begin{figure}%[htb]
 \centerline{\includegraphics[width=8.5cm, height=5.5cm]{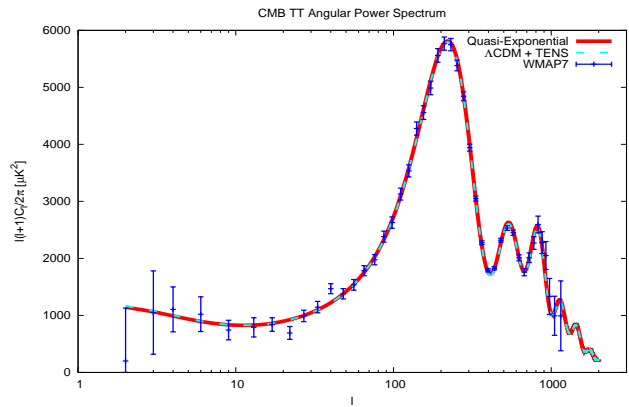}}
 \caption{\label{tt} Variation of CMB angular power spectrum $C_l^{TT}$ for quasi-exponential inflation, the best fit spectra of WMAP7 for $\Lambda$CDM + TENS  and WMAP7 data 
with the multipoles $l$}
\end{figure}
In Fig.\ref{tt} we confront CAMB output of CMB angular power
spectrum $C_l^{TT}$ for quasi-exponential inflation with WMAP seven years data and the best fit spectra of WMAP7 for $\Lambda$CDM + TENS. On
the large angular scales i.e. for low $l$, CMB anisotropy spectrum
is dominated by the fluctuations in the gravitational potential
leading to {\it Sachs-Wolfe} effect.
 From Fig.\ref{tt} we see that the Sachs-Wolfe plateau obtained from quasi-exponential inflation is almost flat confirming a nearly scale invariant spectrum
and resonating with the small spectral tilt $1-n_s=0.0254$ as
estimated from our analysis.

For larger $l$, CMB anisotropy spectrum is dominated by the
acoustic oscillations of the baryon-photon fluid giving rise to
several peaks and troughs in the spectrum.
The heights of the peaks are very susceptible to the baryon
fraction. Also the peak positions are sensitive to the curvature
of the space and on the rate of cosmological expansion hence on
the dark energy and other forms of the matter. In Fig.\ref{tt} the
first and most prominent peak arises at $l=220$ at a height of
$5811{\rm \mu K^2}$ followed by two equal height peaks at $l=537$
and $l=815$. This is in excellent agreement with WMAP seven years
data \cite{wmap} for $\Lambda$CDM background. The direct comparison of our
prediction for $p=0.60$ in Fig.\ref{tt} shows fine match with WMAP
data apart from the two outliers at $l=22$ and $l=40$. The above
analysis is for a representative value of the parameter $p=0.6$ witin its allowed range.

\begin{figure}%[htb]
 \centerline{\includegraphics[width=8.5cm, height=5.5cm]{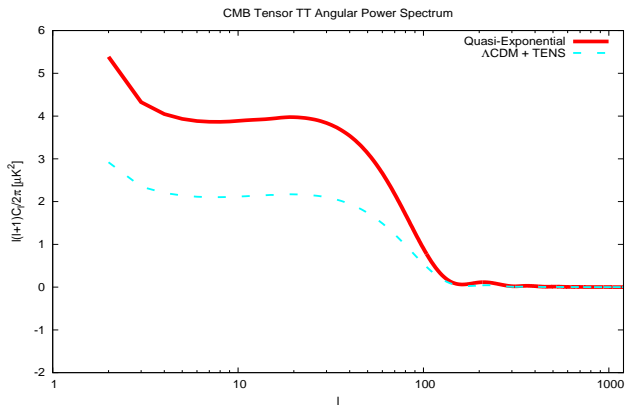}}
  \caption{\label{ttt} Variation of CMB tensor $C_l^{TT}$ angular power spectrum for quasi-exponential inflation and the predicted tensor spectra of WMAP7 for $\Lambda$CDM + TENS  
with the multipoles $l$}
\end{figure}
The gravitational waves generated during inflation also remain
constant on {\it super Hubble} scales having small amplitudes. But
as its wavelength becomes smaller than horizon the amplitude
begins to die off very rapidly. So the small scale modes have no
impact in the CMB anisotropy spectrum only the large scale modes
have little contribution and this is obvious from Fig.\ref{ttt}.

\begin{figure}%[htb]
\centerline{\includegraphics[width=8.5cm, height=5.5cm]{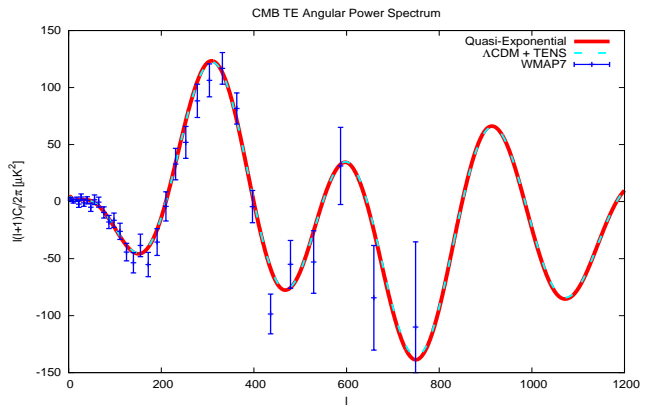}}
\caption{\label{te} Variation of CMB angular power spectrum
$C_l^{TE}$ for quasi-exponential inflation and the best fit spectra of WMAP7 for $\Lambda$CDM + TENS with the multipoles $l$}
\end{figure}

\begin{figure}%[htb]
 \centerline{\includegraphics[width=8.5cm, height=5.5cm]{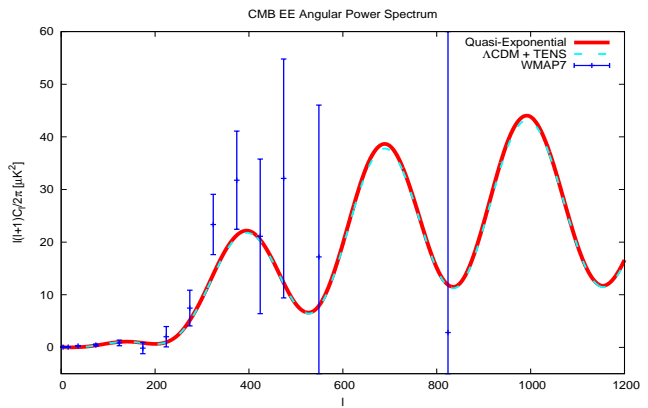}}
  \caption{\label{ee} Variation of CMB angular power spectrum $C_l^{EE}$ for quasi-exponential inflation and the best fit spectra of WMAP7 for $\Lambda$CDM + TENS with the multipoles $l$}
\end{figure}

Further, in Fig.\ref{te} and Fig.\ref{ee} we have plotted CMB TE
and EE angular power spectrum respectively for quasi-exponential inflation and the best fit spectra of WMAP7 for $\Lambda$CDM + TENS 
and compared with
WMAP seven years data. Both the plots resonate fairly well with
the latest WMAP data \cite{wmap}.

Thus, from the entire analysis, it turns out that
quasi-exponential inflation confronts extremely well with WMAP
seven dataset. We expect that the forthcoming data from
PLANCK  would constrain quasi-exponential inflation further.

%%%%%%%%%%%%%%%%%%%%%%%%%%%%%%%%%%%%%%%%%%%%%%%%%%%%%%%%%%%%%%%%%%%%%%%%%%%%%%%%%%%%%%%%%%%%%%%%%%%%%%%%%%%%%%%%%%%%%%%%%%%%%%%%
%%%%%%%%%%%%%%%%%%%%%%%%%%%%%%%%%%%%%%%%%%%%%%%%%%%%%%%%%%%%%%%%%%%%%%%%%%%%%%%%%%%%%%%%%%%%%%%%%%%%%%%%%%%%%%%%%%%%%%%%%%%%%%%%
\section{conclusion}

 In this article we have confronted quasi-exponential models of inflation with WMAP seven year dataset using Hamilton-Jacobi formalism. We have first developed the formalism with a 
phenomenological Hubble parameter and have demonstrated how and to what extent the scenario measures
deviation from standard  exponential inflation ({\it e.g.}, de-Sitter model). The deviation, incorporated through
a new parameter $p$, has then been constrained by estimating the major observable parameters from the model and confronting them
with WMAP seven year dataset. The observable parameters, as obtained from our analysis, turn out to be in good
fit with the latest WMAP data.

 We have further utilized the publicly available code CAMB \cite{camb}
to compare with WMAP seven data \cite{wmap} the matter power spectrum as well
as CMB angular power spectra for the TT, TE and EE modes obtained
from our analysis. This allows us to put stringent  constrains on
the model parameters. CAMB outputs for a emblematic value of the parameter $p$ within its allowed
range $0.586<p<\sqrt{2}$ are in excellent agreement with the latest WMAP data. Values of the most
significant cosmological parameters have also been calculated
using CAMB and have found to fair well with the observational
bounds as given by WMAP seven years data. This leads us
to conclude that quasi-exponential inflation confronts extremely
well with WMAP seven within a certain parameter space.

Nevertheless, another appealing aspect of our analysis is the
possibility of verifying quasi-exponential inflation  by
PLANCK. The current observational bound on the ratio of tensor to scalar amplitudes as given by WMAP is $r<0.36$ at 95\% C.L. \cite{komatsu}. With the ongoing satellite
 experiment PLANCK
 promising to detect it down to the order of $10^{-2}$ \cite{colombo} in near future, this parameter is now playing a pivotal role (along with
non-gaussianity) to discriminate among different models. The
numerical estimation  reveals that the parameter $r$ is of the
order of $10^{-2}$ for quasi-exponential inflation. Thus, the
forthcoming data  from PLANCK can be confronted
directly with quasi-exponential inflation. We are looking forward
to confront this type of inflation with PLANCK and to have more
exciting results in near future.  We also plan to engage ourselves
in estimating cosmological parameters  using another highly useful
code COSMOMC \cite{cosmomc}. Results in this direction will be
reported shortly.

%%%%%%%%%%%%%%%%%%%%%%%%%%%%%%%%%%%%%%%%%%%%%%%%%%%%%%%%%%%%%%%%%%%%%%%%%%%%%%%%%%%%%%%%%%%%%%%%%%%%%%%%%%%%%%%%%%%%%%%%%%%%%%%%%%%
\section*{Acknowledgments}
BKP thanks Sourav Mitra and Dhiraj Hazra for useful discussions.
BKP also thanks Council of Scientific and Industrial Research,
India for financial support through Senior Research Fellowship
(Grant No. 09/093 (0119)/2009). SP is supported by a research
grant from Alexander von Humboldt Foundation, Germany, and  is
partially supported by the SFB-Tansregio TR33 ``The Dark
Universe'' (Deutsche Forschungsgemeinschaft) and the European
Union 7th network program ``Unification in the LHC era''
(PITN-GA-2009-237920).
%\end{document}

%%%%%%%%%%%%%%%%%%%%%%%%%%%%%%%%%%%%%%%%%%%%%%%%%%%%%%%%%%%%%%%%%%%%%%%%%%%%%%%%%%%%%%%%%%%%%%%%%%%%%%%%%%%%%%%%%%%%%%%%%%%%%%
%%%%%%%%%%%%%%%%%%%%%%%%%%%%%%%%%%%%%%%%%%%%%%%%%%%%%%%%%%%%%%%%%%%%%%%%%%%%%%%%%%%%%%%%%%%%%%%%%%%%%%%%%%%%%%%%%%%%%%%%%%%%%%%
%\section*{references}

\end{document}